\documentclass[]{spie} 

\usepackage{amsmath,amsfonts,amssymb}
\usepackage{graphicx}
\usepackage{booktabs}   
\usepackage{siunitx}    
\DeclareSIUnit{\GSPS}{GSPS}
\DeclareSIUnit{\MSPS}{MSPS}
\DeclareSIUnit{\Gbps}{Gbps}
\DeclareSIUnit{\Mbps}{Mbps}
\DeclareSIUnit{\kbps}{kbps}
\DeclareSIUnit{\TB}{TB}
\usepackage[colorlinks=true, allcolors=blue]{hyperref}

\graphicspath{{./}} 

\title{An RFSoC-based Backend and Timing System for the Balloon-borne Very Long Baseline Interferometry Experiment}

\author[a]{Mayukh Bagchi}
\author[a]{Felix M. Thiel}
\author[a]{Laura M. Fissel}
\author[a]{Maggie Oxford}
\author[b,c]{Lindy Blackburn}
\author[a]{Rafael Costa}
\author[d]{Vincent L. Fish}
\author[e,f]{Daryl Haggard}
\author[c]{Michael D. Johnson}
\author[b,c]{Dominic W. Pesce}
\author[d]{Ganesh Rajagopalan}
\author[g]{Javier L. Romualdez}
\author[a]{Aarchi Shah}
\author[h]{Adrian K. Sinclair}
\author[a]{Stephanie St-Jean}
\author[a]{Terry Yang}

\affil[a]{Queen's University, Kingston, ON, Canada}
\affil[b]{Harvard University, Cambridge, MA, USA}
\affil[c]{Center for Astrophysics \textbar{} Harvard \& Smithsonian, Cambridge, MA, USA}
\affil[d]{MIT Haystack Observatory, Westford, MA, USA}
\affil[e]{McGill University, Montr\'eal, QC, Canada}
\affil[f]{Trottier Space Institute at McGill, Montr\'eal, QC, Canada}
\affil[g]{StarSpec Technologies Inc., Canada}
\affil[h]{Johns Hopkins University, Baltimore, MD, USA}

\authorinfo{Further author information: (Send correspondence to M.B.)\\M.B.: E-mail: mayukh.bagchi@queensu.ca}

\newcommand{\spiecid}{1415339}       
\newif\ifspienotice
\spienoticetrue                      

\begin{document}
\maketitle

\ifspienotice
\footnotetext[0]{Copyright 2026 Society of Photo-Optical Instrumentation Engineers
(SPIE). One print or electronic copy may be made for personal use only. Systematic
reproduction and distribution, duplication of any material in this publication for a
fee or for commercial purposes, and modification of the contents of the publication
are prohibited.\par
Published as: M.~Bagchi et~al., ``An RFSoC-based backend and timing system for the
balloon-borne very-long baseline interferometry experiment,'' Proc.\ SPIE
\textbf{14153}, \emph{Radio Telescopes, Technologies, and Methods}, \spiecid\
(17 August 2026), \url{https://doi.org/10.1117/12.3101053}. This is the
author-prepared version, posted in accordance with SPIE's article-sharing policy.}
\fi

\begin{abstract}
We present the design and performance characterization of the digital backend and precision-timing
system for the Balloon-borne Very Long Baseline Interferometry Experiment (BVEX), a pathfinder for
high-frequency stratospheric VLBI at \SI{22}{GHz}. The backend uses one of the four 14-bit
analog-to-digital converter inputs on an AMD-Xilinx RFSoC 4x2. Although the converters support
sampling rates up to \SI{5}{\GSPS}, the flight configuration digitizes the \SIrange{2}{4}{GHz}
intermediate frequency at \SI{4.096}{\GSPS}. \texttt{CASPER} firmware provides both a high-resolution
spectrometer for pointing and receiver verification, and a VLBI acquisition chain with two-bit requantization that records at a rate of about
\SI{8.2}{\Gbps}. The timestamped data packets are sent over 100 Gigabit Ethernet (GbE) to a \SI{16}{\TB} NVMe array in a
storage computer that draws approximately \SIrange{70}{80}{W}. The timing chain uses a Rakon
oven-controlled crystal oscillator as a timing reference while a time-interval counter measures its drift relative to
a GPS reference with approximately \SI{60}{ps} resolution. This is the first deployment of an
RFSoC-based VLBI backend and precision-timing system on a stratospheric balloon. Ground tests
validated the backend, spectrometer, and timing chain. The August 2025 CSA
STRATOS flight ended before reaching the target float altitude because of a balloon failure, and as a result no science
observations were obtained. For the planned 2027 reflight, we are developing a conduction-cooled
data storage computer with \SI{24}{\TB} of NVMe capacity and a direct data path from the 100~GbE
interface to the NVMe array.
\end{abstract}




\section{INTRODUCTION}
\label{sec:intro}

Very long baseline interferometry (VLBI) achieves the finest angular resolution in astronomy by
combining signals recorded at radio telescopes separated by thousands of kilometers. The resolution can be improved by increasing the maximum separation between telescopes (baseline length) or increasing the frequency of light observed. The Event Horizon Telescope, for example,
resolved the shadow of the supermassive black hole in M87 at \SI{230}{GHz} \cite{eht2019m87}. A
stratospheric VLBI station can extend an array beyond the fixed sites available on the ground (thereby improving the $uv$ coverage) and can
observe above most atmospheric water vapor (allowing observations at frequencies that are not accessible from most ground-based sites). The balloon platform, however, imposes strict limits on
mass, power, and thermal control, and additionally the gondola motion will change the phase of the measured wavefront.

Precise timing is fundamental to VLBI as each telescope records its signal independently and the
signals are combined only afterward in a correlator. Thus, every station needs a stable, accurate clock. Hydrogen masers, which are
commonly used as timing standards at VLBI observatories, exceed the mass and power
budget of a balloon payload. An alternative strategy is to use a compact oven-controlled crystal oscillator (OCXO) for short-term
frequency stability and measure its drift against GPS to reference the recorded data to Coordinated
Universal Time (UTC). 

The Balloon-borne Very Long Baseline Interferometry Experiment (BVEX) is a \SI{22}{GHz} pathfinder
for a high-frequency VLBI station. The payload includes a 0.9 m Cassegrain radio telescope and receiver, mounted on a pointed aluminum frame gondola. The signal is digitized and recorded on board for later
correlation with ground stations. The
thermal noise of a VLBI baseline scales with the geometric mean of the two stations'
system-equivalent flux densities (SEFD). Pairing the small BVEX telescope with large ground-based dishes therefore
reduces the baseline noise and preserves enough sensitivity to detect fringes on bright sources. The telescope, receiver, pointing system and the overall experiment are described in a companion paper \cite{thiel2026bvex} and \cite{BVEXInstrument}. The receiver covers roughly
\SIrange{21}{23}{GHz}, a band that contains the \SI{22.235}{GHz} water-maser line, a strong line
source for pointing checks that can be used for single-dish test observations of the BVEX telescope. The same receiver supports the broadband VLBI observations that are the primary science mode. Doi et al. developed a similar balloon-borne VLBI platform (at approximately
\SI{20}{GHz}), but reported a systems-design feasibility study rather than an actual flight
\cite{doi2019balloon}.

The primary goal of the first flight was to detect interferometric fringes on the bright AGN 3C\,84 and 3C\,454.3 while simultaneously observing with the Very Long Baseline Array (VLBA), the
Effelsberg \SI{100}{m} telescope, and the MIT Haystack Observatory \SI{37}{m} telescope. To correlate
with these telescopes, our BVEX backend must match the array specifications. In VLBI mode we therefore
sample at \SI{4.096}{\GSPS}, enough to Nyquist-sample the full \SI{2}{GHz} receiver band. Recording
the 14-bit samples would produce about \SI{57.3}{\Gbps}, so we follow standard VLBI practice and
quantize to two bits, capping the data rate at \SI{8.2}{\Gbps} while retaining most of the
sensitivity \cite{tms2017interferometry}. Recording at that rate for hours requires fast networking and terabytes
of storage.

This paper describes the backend and timing system that meet these requirements and reports their
ground validation and flight behavior. The backend runs on a single RFSoC 4x2 board, a
radio-frequency system-on-chip (RFSoC) programmed with the open-source tool flow of the Collaboration for Astronomy Signal
Processing and Electronics Research (\texttt{CASPER}) \cite{hickish2016casper}. The timing system pairs an
OCXO with GPS-referenced drift monitoring. BVEX was launched on August 29, 2025, from the Canadian Space Agency (CSA)
STRATOS base in Timmins, Ontario; a balloon failure ended the flight before it reached float
altitude, so no science observations were obtained. We are rebuilding the instrument for a planned
2027 reflight.

The paper is organized as follows: Section~\ref{sec:backend} presents the backend architecture and
Section~\ref{sec:timing} the precision-timing system. Section~\ref{sec:integration} covers system
integration and the operational software, Section~\ref{sec:results} reports ground testing and the system performance during the August 2025 flight, and Section~\ref{sec:bvex2} discusses the lessons learned and our plans for a
next-generation backend for the 2027 reflight. Conclusions are presented in Section~\ref{sec:conclusions}.

\section{BACKEND SYSTEM ARCHITECTURE}
\label{sec:backend}

\begin{figure}[tb]
  \centering
  \includegraphics[width=\textwidth]{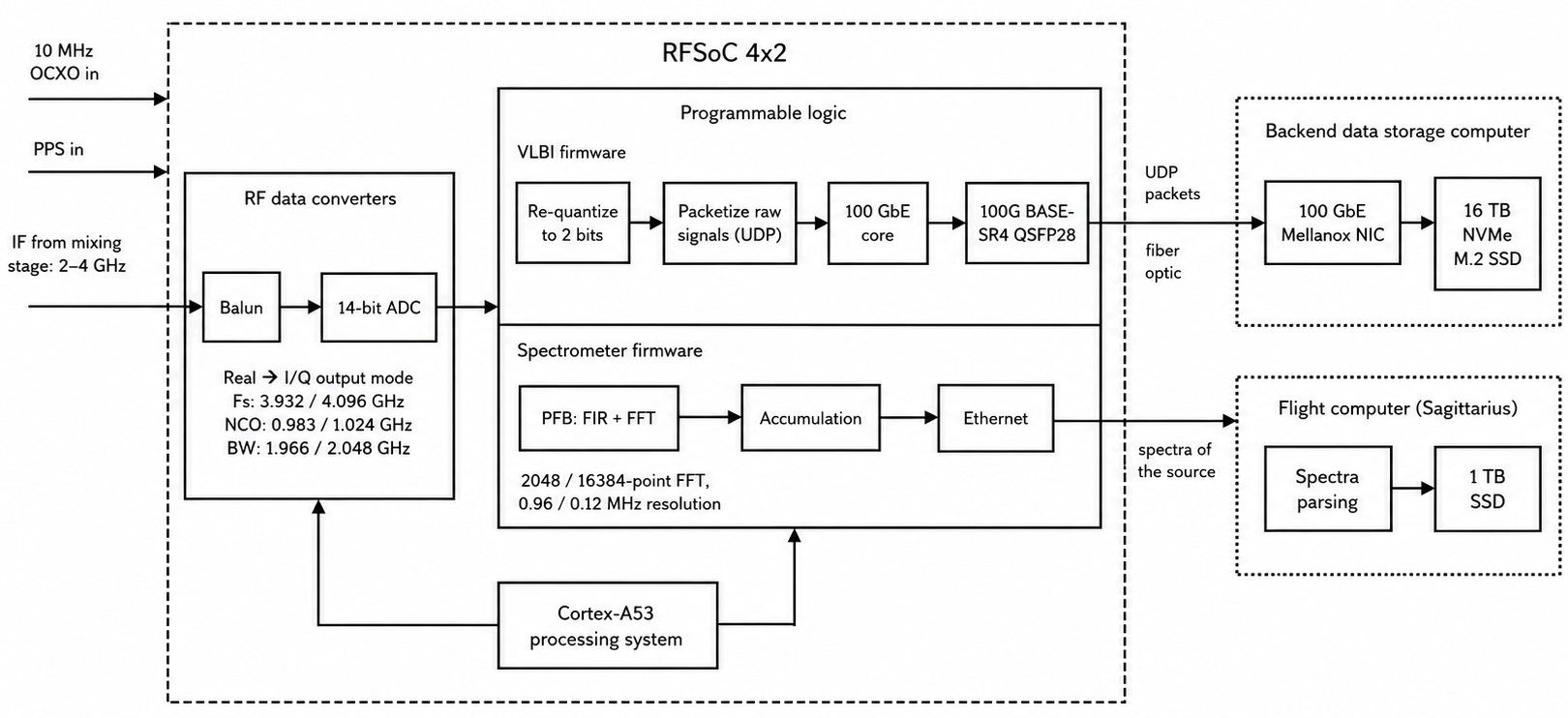}
  \caption{End-to-end architecture of the BVEX digital backend. A single RFSoC 4x2 digitizes the
  receiver IF and, depending on the loaded firmware, produces either accumulated spectra for
  pointing and receiver verification or a two-bit baseband stream with a custom timing header that is
  recorded over 100~GbE. The data converters and the timestamp generator are driven by the
  \SI{10}{MHz} reference and the 1~PPS signal from the precision-timing chain
  (Section~\ref{sec:timing}).}
  \label{fig:backend-block}
\end{figure}

The backend digitizes the receiver intermediate frequency (IF), described in the companion paper \cite{thiel2026bvex}, on the RFSoC, processes it in
firmware, and delivers one of two products: power spectra for pointing and telescope performance tests,
and a two-bit baseband stream for VLBI observations. Both products share a common front end (on-chip
digitization and digital down-conversion) before splitting into mode-specific processing. The two
modes are built as separate firmware bitstreams which are then deployed on the field-programmable gate array (FPGA) during flight depending on the observing mode used. The amplitude spectral data are read out by the flight computer at a
slow rate, while the VLBI data stream is packetized with timing counters and sent over 100 Gigabit Ethernet
(100~GbE) to a dedicated storage computer. Figure~\ref{fig:backend-block} shows the complete signal flow. This section
follows that data flow, from the hardware platform through to storage.

\subsection{RFSoC 4x2 platform}
\label{subsec:rfsoc}

The BVEX backend is built on a single Real Digital RFSoC 4x2 board \cite{realdigitalrfsoc4x2}, whose
core is the AMD-Xilinx Zynq UltraScale+ RFSoC XCZU48DR, a third-generation device that integrates
wideband data converters, programmable logic (PL), and a quad-core Arm Cortex-A53 processing
system (PS) on one chip \cite{amd2023rfsoc}. The board uses four 14-bit analog-to-digital
converter (ADC) channels that run at up to \SI{5}{\GSPS} (giga-samples per second), each with an
on-chip digital down-converter that includes a numerically controlled oscillator (NCO) and
selectable decimation. The PL fabric carries the custom firmware, and the device exposes
high-speed transceivers terminated in a QSFP28 cage that we use for the 100~GbE output. The BVEX receiver output is single-polarization and single-sideband, so BVEX uses only one ADC channel of the board.

We discipline the data converters from the BVEX timing chain rather than from the board's internal VCXO 
reference. The \SI{10}{MHz} output of the OCXO enters the board's external clock input, an SMA
connector that feeds the on-board clocking network. A Texas Instruments LMK04828 jitter
cleaner locks to the external reference, generates the programmable-logic reference and drives
the LMX2594 RF synthesizers which produce the converter sample clocks. The LMK and LMX devices
are programmed at run time from the board's processing system, using prepared register files that
select the external reference. A 1~PPS (one pulse-per-second) input from the GPS provides the synchronization to
UTC, so the backend digitization is referenced to the timing system described in
Section~\ref{sec:timing}. The firmware fabric runs at a \SI{256}{MHz} system clock. The board runs
a Linux-based \texttt{CASPER} software stack on its PS and is housed in a custom enclosure with passive
thermal management; in operation the RFSoC dissipates roughly \SIrange{25}{35}{W}.

\subsection{CASPER firmware and tool flow}
\label{subsec:casper}

The BVEX RFSoC firmware\footnote{\url{https://github.com/mayukh4/BVEX_rfsoc_4x2}} was developed using the open-source \texttt{CASPER} tool flow \cite{hickish2016casper}. Signal
processing is assembled in \texttt{MATLAB} and \texttt{Simulink} from the \texttt{CASPER} digital signal processing (DSP)
library (\texttt{mlib\_devel}), compiled to an FPGA bitstream through Xilinx
\texttt{Vivado}, and loaded and controlled over a network link from the
flight computer using the \texttt{casperfpga} library. This modular nature of the toolflow coupled with open source \texttt{CASPER} tutorial designs allowed us to build the firmware design from
reusable blocks (the RF data converter (RFDC) interface, polyphase filterbank, Fourier transform, vector accumulators,
and Ethernet cores) and iterate the spectrometer and VLBI designs quickly.

The spectrometer and the VLBI acquisition chain are implemented as separate bitstreams, and only
one resides on the device at a time. Before an observation the BVEX Control Program (\texttt{BCP}) \footnote{\url{https://github.com/fissellab/bcp}}, the
flight software that commands the payload, selects the appropriate bitstream and programs it onto
the RFSoC over the command link. Observing mode selection is therefore an operational choice made by the user.

\subsection{Digitization and digital down-conversion}
\label{subsec:digitization}

The receiver delivers a \SIrange{2}{4}{GHz} IF to an SMA input on the board. We sample this band
directly in the second Nyquist zone of the ADC, which removes the need for a final analog
down-conversion stage. Inside the RFDC, the NCO mixes the band of interest to complex baseband and
a factor-of-two decimation sets the output bandwidth, so that each observing mode is defined by its
sample rate and NCO setting.

In VLBI mode the ADC runs at \SI{4.096}{\GSPS} with the NCO at \SI{1024}{MHz} and decimation by
two, producing complex samples at \SI{2048}{\MSPS} (mega-samples per second) across a
\SIrange{0}{2048}{MHz} baseband. In
spectrometer mode the ADC runs at \SI{3932.16}{\MSPS} with the NCO at \SI{983.04}{MHz} and
decimation by two, spanning \SIrange{0}{1966.08}{MHz}. The 14-bit ADC samples are carried as
16-bit signed words through the DSP fabric.

\subsection{Spectrometer for pointing and receiver verification}
\label{subsec:spectrometer}

\begin{figure}[tb]
  \centering
  \includegraphics[width=\textwidth]{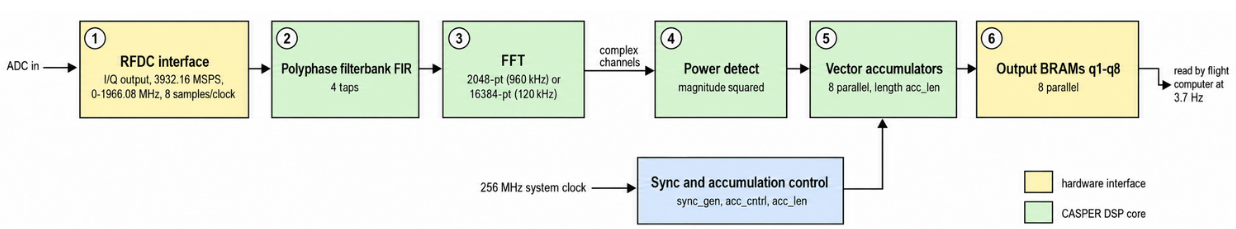}
  \caption{The spectrometer firmware showing the simplified signal flow, with the stages numbered:
  the RFDC interface, which delivers complex baseband at \SI{3932.16}{\MSPS} across
  \SIrange{0}{1966.08}{MHz} at eight samples per clock (1); a four-tap polyphase filterbank FIR (2)
  and FFT (3) that channelize the band, in the 2048-point coarse mode or the 16384-point fine mode;
  the power detector that squares the complex channels to form a power spectrum (4); eight parallel
  vector accumulators (5); and the eight output BRAMs (q1 through q8) that the flight computer reads
  at about \SI{3.7}{Hz} (6). A separate control block sets the accumulation length and sync from
  software registers. Every stage is a stock \texttt{CASPER} library block; the same tool flow builds the
  separate VLBI acquisition bitstream of Figure~\ref{fig:vlbi-firmware}.}
  \label{fig:spectrometer-firmware}
\end{figure}

While the spectrometer is not the primary science observing mode for BVEX, it is important for testing the telescope, verifying the performance of the instrument, and, by pointing at bright water maser sources, verifying that the telescope pointing control system works and is able to track bright sources. The spectrometer firmware, shown in Figure~\ref{fig:spectrometer-firmware}, channelizes
the complex baseband with a four-tap polyphase filterbank (PFB) followed by a fast Fourier transform
(FFT), and accumulates the resulting power spectrum in block RAM (BRAM) \cite{price2021pfb}. The
flight computer reads the accumulated spectrum through the \texttt{casperfpga} interface at about
\SI{3.7}{Hz}. To sustain throughput, the transform output is written to eight BRAMs (q1 through q8)
that are read in parallel and interleaved in software.

We built two resolution modes as separate bitstreams. The coarse mode uses a 2048-point FFT, giving
channels of about \SI{960}{kHz} across the \SI{1966.08}{MHz} band; it is used on bright sources such
as the Moon and the Sun to confirm that the receiver and the data acquisition chain are working. The fine
mode uses a 16384-point FFT, giving channels of about \SI{120}{kHz}, and was built specifically for observations of the \SI{22.235}{GHz} water maser, whose typical spectral line width of a few tens of kHz is unresolved by the coarse
mode. Lengthening the transform from 2048 to 16384 points required deepening each of the eight
parallel BRAMs to 2048 words and retuning the sync generator period and the accumulation length to
match the longer pipeline. For telemetry from the balloon telescope to the ground, the flight computer extracts a 167-channel
window (about \SI{20}{MHz} wide) centered on \SI{22.235}{GHz}, subtracts a median baseline, and
transmits the reduced spectrum; the full spectra are stored on board. The performance of the spectrometer in ground-based test observations is discussed in Section~\ref{subsec:w49n-detection}.

\subsection{Two-bit requantization}
\label{subsec:requant}

For the VLBI firmware the wideband complex baseband must be reduced to two bits per sample before recording. We
use a fixed-threshold, four-level (two-bit) quantizer: each in-phase and quadrature sample is mapped
to one of four levels by comparing it against thresholds at zero and at $\pm1000$ in the 16-bit
datapath units. This is the standard two-bit VLBI quantization, which retains close to 88\% of the
signal-to-noise ratio of continuous sampling when the thresholds are placed near $\pm1\sigma$ of the
noise distribution \cite{tms2017interferometry}.

Our thresholds are fixed rather than adaptive, so they are optimal at only one input power level. We
set them for the expected noise level at the quantizer input; departures from that level, for
example from receiver gain drift or changing sky and source conditions, move the operating point
away from the optimum and incur a modest sensitivity penalty. We adopted fixed thresholds for
firmware simplicity and deterministic behavior on the first flight, and note that power-tracking
adaptive thresholds are a planned upgrade for the 2027 reflight. On each \SI{256}{MHz} clock cycle the
requantizer processes eight complex samples, reducing a 256-bit word (eight complex samples at 16
bits per component) to 32 bits (eight at two bits per component).

\subsection{VLBI Data Packing and Packet Timestamping}
\label{subsec:vdif}

Before transmission, a packer groups eight consecutive 32-bit quantized words into a single 256-bit
word, an 8:1 aggregation that fills the 100~GbE payload efficiently. Each complex sample carries two
bits in-phase and two bits quadrature, so at \SI{2048}{\MSPS} the recorded data rate is \SI{8.192}{\Gbps}
(the \SI{4.096}{\GSPS} sampling at two bits per sample), which we quote in this work as
\SI{8.2}{\Gbps}.

\begin{figure}[tb]
  \centering
  \includegraphics[width=\textwidth]{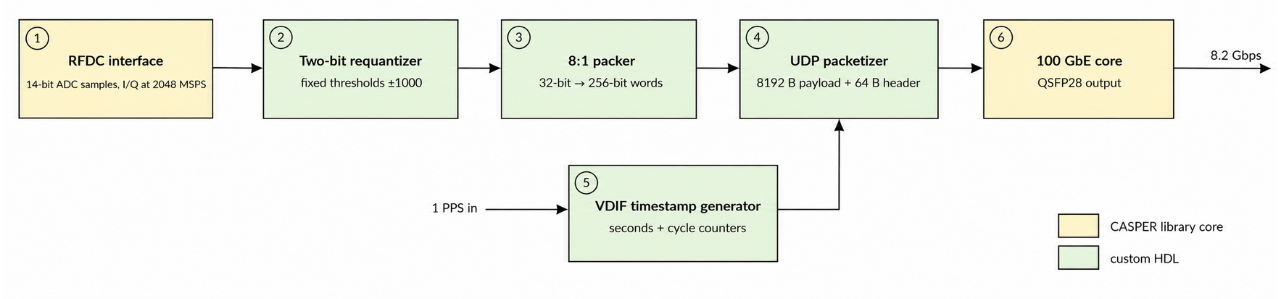}
  \caption{The VLBI acquisition firmware showing the corresponding simplified signal flow, with the stages numbered:
  the RFDC interface (1), the custom HDL two-bit
  fixed-threshold requantizer (2) and 8:1 packer (3) (Section~\ref{subsec:requant}), the UDP
  packetizer that assembles the 8192-byte payload and 64-byte header (4), the timestamp generator
  that writes the seconds and cycle counts used for offline VDIF conversion (5), and the 100~GbE core that drives
  the QSFP28 output (6). The spectrometer firmware is a separate bitstream built in the same tool
  flow.}
  \label{fig:vlbi-firmware}
\end{figure}

Each packet carries two timing counters intended for conversion to the VLBI Data Interchange Format
(VDIF) in post-processing \cite{whitney2009vdif}. Standard VDIF packets store a six-month reference-epoch
index, seconds from that epoch, and a frame number within the second. The BVEX firmware instead
writes a 32-bit seconds counter and a 32-bit \SI{256}{MHz} cycle counter into a custom 64-byte
packet header. A GPS PPS increments the seconds counter, while the OCXO-derived system clock
drives the cycle counter. Figure~\ref{fig:vlbi-firmware} shows the \texttt{Simulink} top level of this chain,
with the custom HDL blocks alongside the \texttt{CASPER} library cores.

\begin{figure}[tb]
  \centering
  \includegraphics[width=0.88\textwidth]{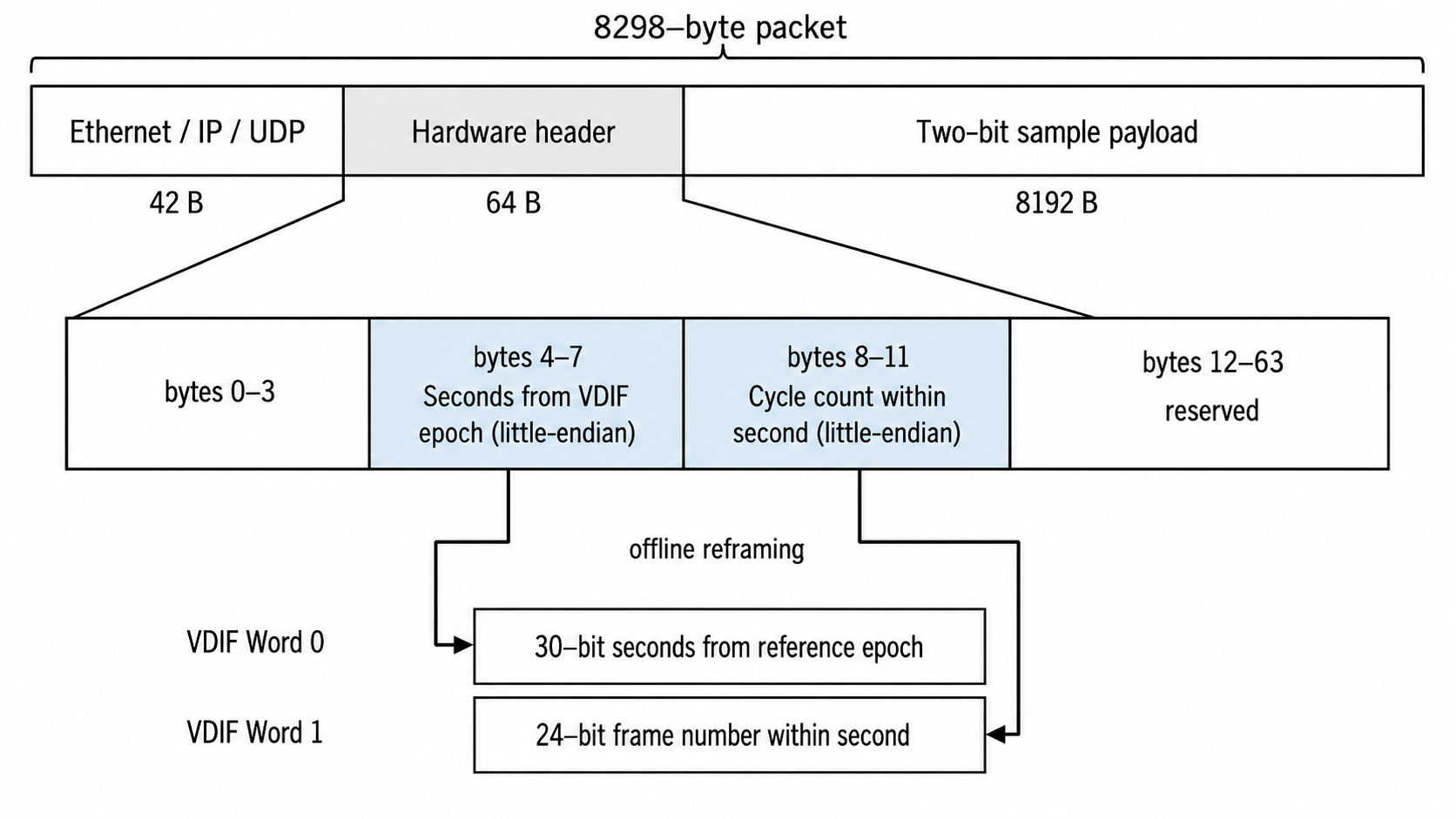}
  \caption{Structure of the 8298-byte VLBI packet and the intended offline conversion of its custom
  timing fields to VDIF. The 2025 hardware header stored a 32-bit seconds counter and a 32-bit clock
  cycle counter; these were not complete VDIF Word~0 and Word~1 fields in the streamed packet.}
  \label{fig:packet}
\end{figure}

The resulting packet is 8298 bytes: 42 bytes of Ethernet, IP, and UDP headers, a 64-byte hardware
header, and an 8192-byte data packet. Figure~\ref{fig:packet} summarizes the packet
layout and the intended offline mapping of these counters into VDIF fields.

The 2025 firmware implemented the on-board timestamp generator where each packet carried a seconds count
incremented by the GPS PPS and a clock-cycle count from the OCXO-derived system clock, in an intermediate format designed for
post-flight conversion to VDIF. We plan to make the packets more compatible with VDIF as discussed in Section~\ref{subsec:bvex2-firmware}.

\subsection{Data path, compute, and storage}
\label{subsec:storage}

\begin{figure}[tb]
  \centering
  \includegraphics[width=\textwidth]{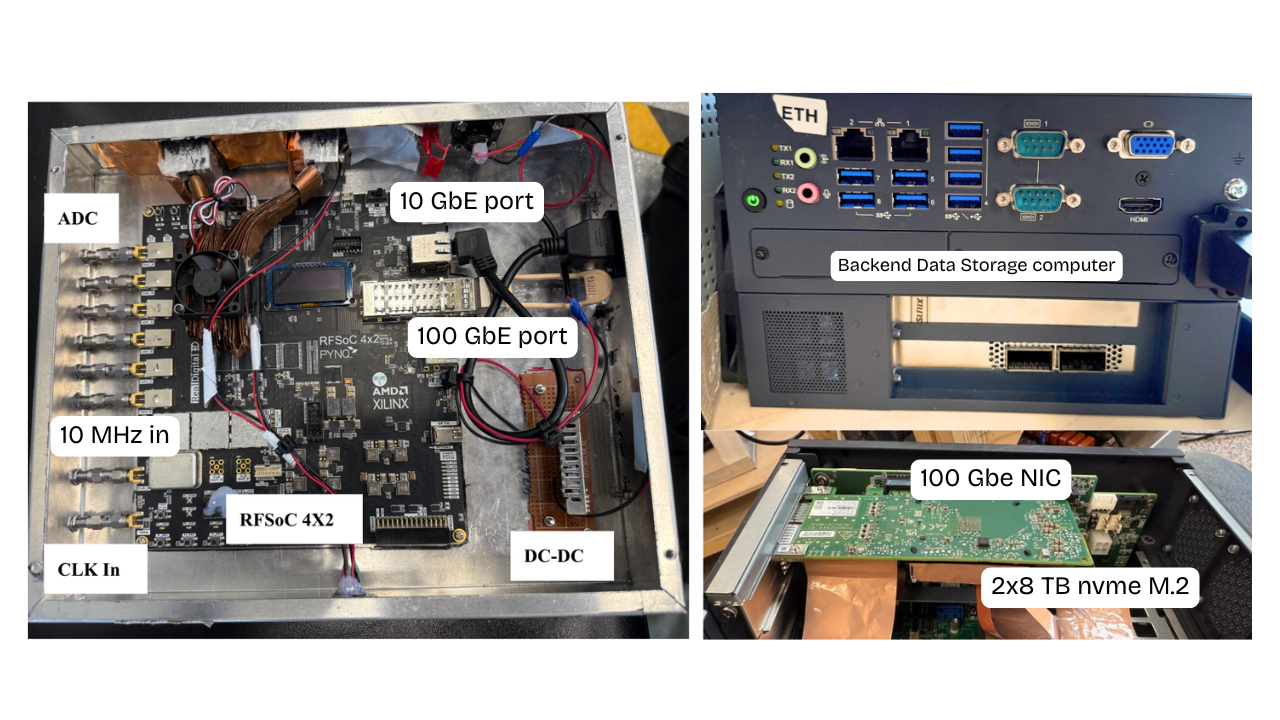}
  \caption{The BVEX backend hardware. Left: the RFSoC 4x2 in its flight enclosure, with the ADC
  input from the receiver, the \SI{10}{MHz} reference connection to the board's external clock input, the DC-DC power
  stage, and the 100~GbE QSFP28 output. Right: the backend data storage computer, with the
  ConnectX-5 100~GbE NIC and the two \SI{8}{\TB} NVMe drives. The OCXO frequency standard and the
  time-interval counter are shown in Figure~\ref{fig:timing-hw}.}
  \label{fig:hardware}
\end{figure}

The two-bit VLBI data stream leaves the RFSoC on the QSFP28 port as 100~GbE and is captured by a
dedicated backend computer. For the 2025 BVEX flight we flew an Advantech MIC-770V3 computer
fitted with a Mellanox ConnectX-5 100~GbE network interface card (NIC) and two \SI{8}{\TB} NVMe M.2
solid-state drives, for a total of \SI{16}{\TB} of storage. With a VLBI data rate of about \SI{8.2}{\Gbps}, this provides roughly \SI{4.3}{\hour} of continuous recording. The backend computer draws
approximately \SIrange{70}{80}{W} of power during nominal operations. The capture
and logging software, and the integration of this computer with the rest of the payload, are
described in Section~\ref{sec:integration}. Figure~\ref{fig:hardware} shows the flight backend
hardware.

\section{PRECISION TIMING SYSTEM}
\label{sec:timing}

\begin{figure}[tb]
  \centering
  \includegraphics[width=\textwidth]{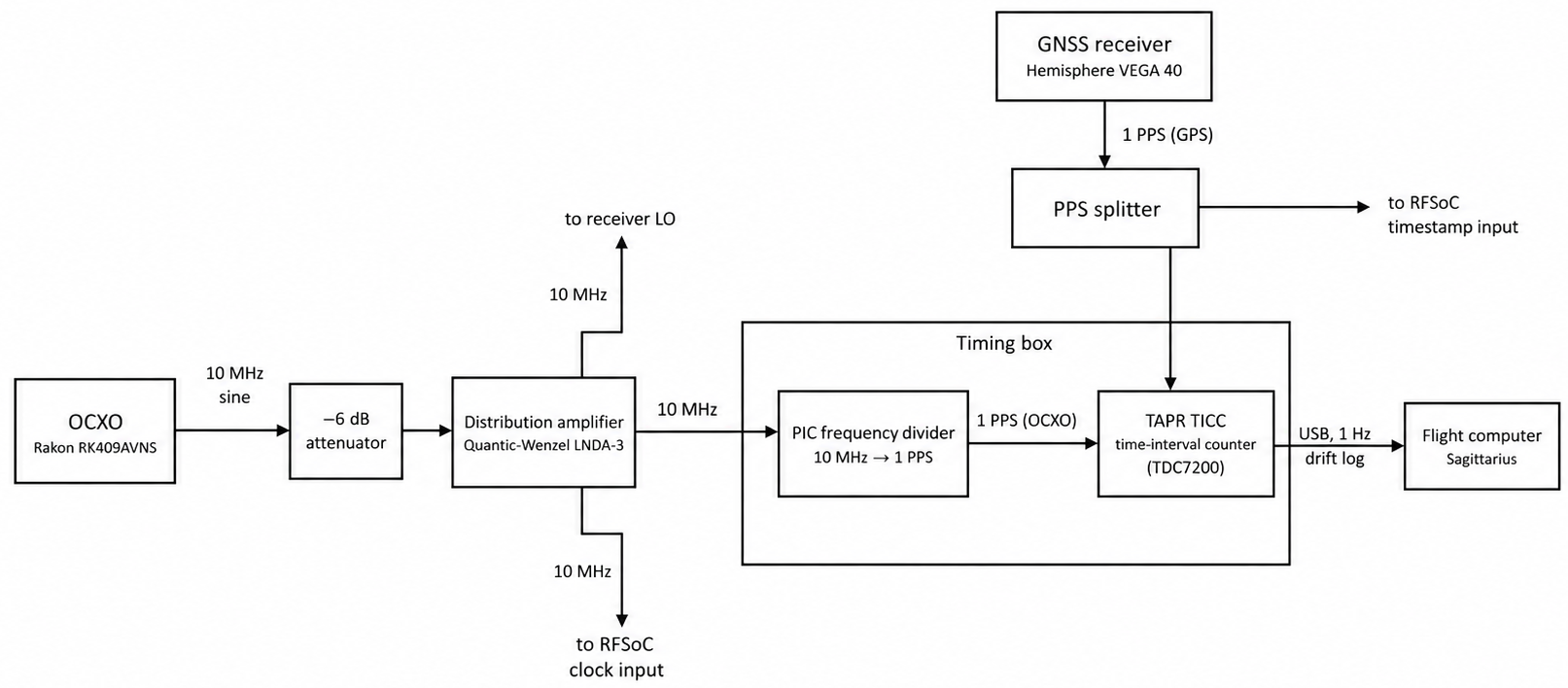}
  \caption{Architecture of the BVEX precision-timing chain. A single \SI{10}{MHz} OCXO reference is
  distributed to the receiver local oscillator, the RFSoC, and a time-interval counter; the counter
  measures the drift of the free-running OCXO against a GPS \SI{1}{Hz} reference once per second,
  building the record used to refer the recorded data to UTC during correlation. The same GPS
  1~PPS drives the VDIF timestamp generator (Section~\ref{subsec:vdif}).}
  \label{fig:timing-block}
\end{figure}

VLBI correlation cross-multiplies the voltage samples recorded at the separate stations and
searches for an interferometric fringe over a range of geometric delays. Therefore typically
every station tags its samples with a high-precision time standard that keeps the phase coherent across the
integration. Ground observatories commonly use hydrogen masers as time standards, which are then compared against GPS time to track their offset from UTC. Such a maser exceeds the BVEX mass and power
budget. For BVEX we instead use a low-phase-noise OCXO, which meets the short-term stability requirement over a
one-second VLBI coherence time in a compact, low-power package. Figure~\ref{fig:timing-block} shows a schematic of the
timing chain. As a part of the timing system the oscillator's drift against GPS is
monitored continuously, so the recorded timestamps can be referred to UTC in post-processing. 

\subsection{Frequency standard and distribution}
\label{subsec:ocxo}

The timing reference and frequency standard is a Rakon RK409AVNS OCXO that outputs a \SI{10}{MHz} sine wave and is powered by a
\SI{12}{V} supply \cite{rakon2022rk409}. The RK409 is a flight-qualified oscillator built for low-Earth-orbit
satellites, which is appropriate for a stratospheric environment, and it dissipates only a few watts of power. In
flight, the OCXO box is housed inside a small sealed pressure vessel held at \SI{25}{\celsius} to within
\SI{0.1}{K} by a thermoelectric controller, so the oscillator operates at the pressure and
temperature for which it is specified. We operate
the OCXO without GPS steering to preserve its intrinsic short-term stability and avoid coupling GPS timing
noise into the \SI{10}{MHz} output through a disciplining loop. Free-running operation introduces a
slow drift relative to UTC, which we measure and correct post-flight (see Section~\ref{subsec:drift}).

A Quantic-Wenzel LNDA-3 low-noise distribution amplifier buffers the \SI{10}{MHz} reference output from the OCXO as shown in Figure~\ref{fig:timing-block} and
splits it three ways \cite{quanticwenzel_lnda3}: to the receiver's phase-locked local oscillator, to
the RFSoC clocking network, and to the drift-monitoring counter as discussed in Section~\ref{subsec:stability}. Driving the receiver and the
digitizer from one reference keeps the whole signal chain coherent. In the backend, the on-board
clocking network locks to this same \SI{10}{MHz} at the board's external clock port and derives the
ADC sample clock from it (Section~\ref{subsec:rfsoc}).

\subsection{Drift monitoring against GPS}
\label{subsec:drift}

A free-running OCXO accumulates timing offset relative to UTC at a rate set by its fractional
frequency error. At our measured rate of $-49.4$\,ns/s, a ten-minute scan accumulates roughly
\SI{30}{\micro\second} of offset. This offset cannot be ignored in the correlator delay model. We
therefore record the instantaneous phase difference
between the OCXO and GPS for the whole flight. With a correctly initialized packet clock, this time
series supplies the oscillator correction needed to place BVEX data on the UTC timescale used by the
ground stations.

\begin{figure}[tb]
  \centering
  \includegraphics[width=\textwidth]{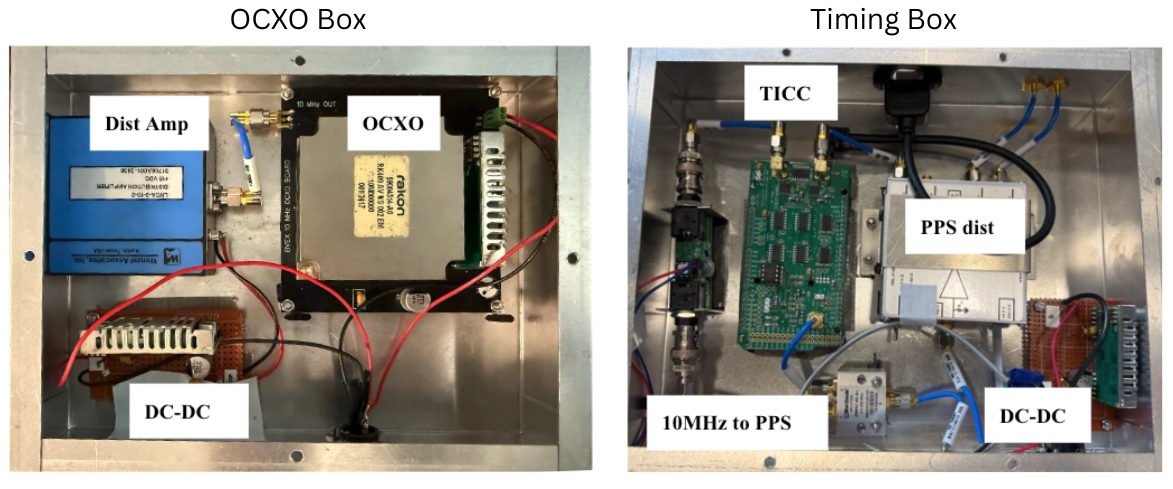}
  \caption{The BVEX timing hardware. Left: the OCXO box, housing the Rakon RK409AVNS frequency
  standard, the LNDA-3 distribution amplifier, and the DC-DC power stage. Right: the timing box,
  which houses the TAPR TICC time-interval counter, the \SI{10}{MHz}-to-1~PPS divider, and the
  1~PPS distribution. This is the detailed view of the timing hardware referenced in
  Figure~\ref{fig:hardware}.}
  \label{fig:timing-hw}
\end{figure}

The comparison is made by a time-interval counter. A 1~PPS signal from a Hemisphere VEGA 40
Global Navigation Satellite System (GNSS) receiver \cite{hemisphere_vega40} provides the UTC
reference edge, and a PIC-microcontroller
divider counts the \SI{10}{MHz} reference down to a \SI{1}{Hz} square wave that represents the OCXO. A
TAPR Time Interval Counter (TICC), built on an Arduino Mega 2560 and a Texas Instruments TDC7200
time-to-digital converter \cite{ti2015tdc7200}, measures the interval between the two edges once per
second and logs the result over USB to the control computer. The TICC has a resolution of about \SI{60}{ps},
with root-mean-square (RMS) jitter below \SI{100}{ps}. A passive splitter sends the same GPS 1~PPS to both the TICC
and the RFSoC timestamp generator, so the packet-counter transitions
(Section~\ref{subsec:vdif}) and the drift record share one reference edge. Figure~\ref{fig:timing-hw}
shows the oscillator and counter hardware.

\subsection{Frequency stability and phase-error budget}
\label{subsec:stability}

We characterized our RK409AVNS at MIT Haystack Observatory against a reference Rakon ULN OCXO. In
three of the four runs the oscillator temperature was stabilized at \SI{25}{\celsius}; for the fourth run the temperature stabilization was turned off. Because this measurement compares two oscillators, the recorded Allan deviation \cite{riley2008stability} combines the contributions of our RK409AVNS and the reference ULN OCXO; across all four runs the recorded value stayed at $\sigma_y(1\,\mathrm{s}) \leq 2.55 \times 10^{-13}$, including the run with the temperature stabilization off. For uncorrelated oscillator noise, the measured Allan variance is the sum of the two oscillators' Allan variances. The largest measured comparison value therefore provides a conservative upper limit of $2.55 \times 10^{-13}$ for our RK409AVNS OCXO. If the two oscillators contribute equally, the corresponding single-oscillator value is $1.80 \times 10^{-13}$. Both values are below the datasheet limit of $1 \times 10^{-12}$. A fractional frequency instability
$\sigma_y(\tau)$ over a coherence time $\tau$ produces an accumulated phase error of order
$2\pi f_{\mathrm{obs}}\,\sigma_y(\tau)\,\tau$ at the observing frequency $f_{\mathrm{obs}}$. Therefore at
\SI{22}{GHz} over a one-second coherence time, the conservative upper limit of $2.55 \times 10^{-13}$ corresponds to a phase error below about $35$\,mrad, well within the $\sim 0.5$\,rad beyond which coherence losses become significant.
This phase error is due to the OCXO alone. The balloon's motion adds a separate phase error, set by how precisely the
gondola position, attitude, and velocity can be reconstructed, which the companion paper treats in detail
\cite{thiel2026bvex}.

\begin{figure}[tb]
  \centering
  \includegraphics[width=0.9\textwidth]{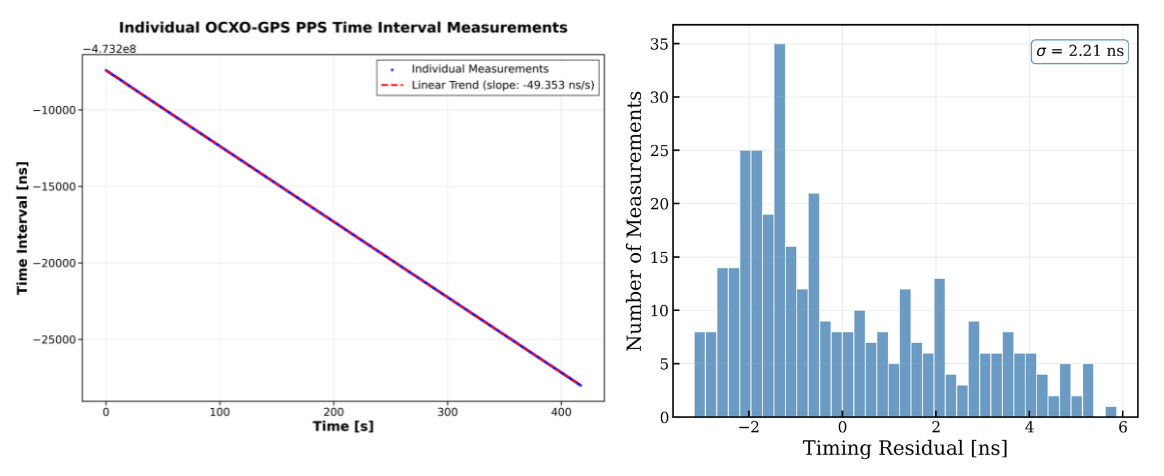}
  \caption{Measured drift of the free-running OCXO against the GPS reference. The interval between
  the GPS 1~PPS and the OCXO-derived \SI{1}{Hz} edge drifts linearly at $-49.4$\,ns/s (a
  fractional frequency offset of $-49.4$\,ppb); the residuals about the linear trend have a standard
  deviation of $2.21$\,ns. The measurement demonstrates that the counter resolves the OCXO drift
  used for post-flight timestamp correction.}
  \label{fig:ocxo-drift}
\end{figure}

The drift rate discussed in Section~\ref{subsec:drift} was measured in a test of the flight timing chain shown in Figure~\ref{fig:ocxo-drift}.
Over
\SI{450}{s} the GPS--OCXO interval fell linearly at $-49.4$\,ns/s, a fractional frequency offset of
$-49.4$\,ppb, and the residuals after removing that linear trend had a standard deviation of
$2.21$\,ns. This residual is dominated by the GPS 1~PPS and the time-interval counter rather than the OCXO; at the one-second sampling it corresponds to a per-sample fractional-frequency scatter of about $2 \times 10^{-9}$, which averages down over the linear fit. The fit demonstrates that the counter measures the slow frequency offset and its long-term drift; the sub-second coherence is set by the OCXO itself (its Allan deviation), which the \SI{1}{Hz} counter cannot resolve. During BVEX flight observations, the drift measured by the time-interval counter can be used to monitor the slow frequency drift of the OCXO \SI{10}{MHz} clock.

\subsection{Balloon gondola vibration environment and phase noise}
\label{subsec:vibration}

A gondola suspended from a stratospheric balloon pendulates continuously due to wind shear and the motion of the flight
train. These low-frequency oscillations couple into the oscillator through its acceleration
sensitivity and can add phase noise that could degrade the correlation with ground-based telescopes. We measured the dynamical environment of the CARMENCITA balloon gondolas used by the Centre National d'\'Etudes Spatiales (CNES) and the CSA by flying BVEXTracker, a small precursor experiment to BVEX, consisting of a box with many different position and attitude tracking sensors that were read out by a Raspberry Pi, which flew as a piggyback on a CNES gondola from Timmins, Ontario \cite{payeur2024realizations}, in 2023 and recorded five to six hours
of three-axis accelerometer data at float. Balloon pendulation concentrates its power at low frequencies, below about \SI{5}{Hz}. The gondola
is most stable during quiescent intervals, which give the best pointing and the lowest oscillator
phase noise, so VLBI observations are scheduled for these periods. We therefore characterize the
acceleration over such an interval: across the measured \SIrange{0.1}{8}{Hz} band the acceleration
spectral densities are about $21$, $15$, and $14\,\mu\mathrm{g}/\sqrt{\mathrm{Hz}}$ on the $a_x$,
$a_y$, and $a_z$ axes, with the largest vibration on the horizontal $a_x$ axis.

\begin{figure}[tb]
  \centering
  \includegraphics[width=0.9\textwidth]{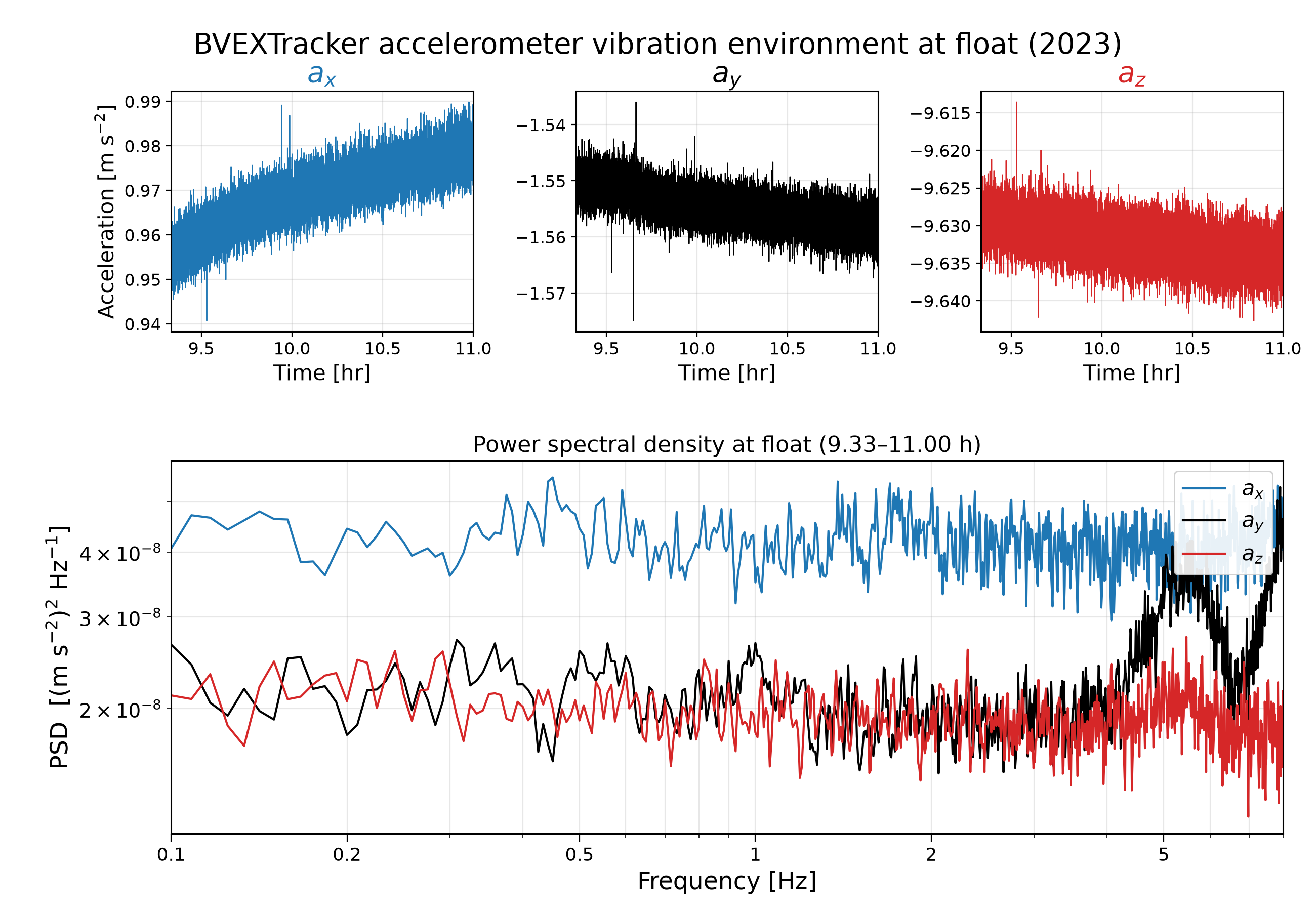}
  \caption{The vibration environment measured by the BVEXTracker precursor payload at float in
  2023. Top: the three-axis accelerometer time series over a quiescent interval at float. Bottom:
  the power spectral density of the same interval, showing the per-axis accelerometer noise floors.
  Folding these accelerations through Equation~\ref{eq:vibration-phase-noise} with
  $\Gamma = 0.8$\,ppb/g gives a vibration-induced phase noise that stays well below the oscillator's
  static datasheet limits across the measured band, so the platform motion does not limit the
  oscillator's phase stability during these intervals.}
  \label{fig:ocxo-vibration}
\end{figure}

A crystal oscillator's fractional frequency shifts in proportion to applied acceleration through its
g-sensitivity $\Gamma$ \cite{filler1988gsensitivity}; for the RK409AVNS $\Gamma = 0.8$\,ppb/g. The
single-sideband phase noise this adds at a vibration frequency $f_v$ is
\begin{equation}
  \mathcal{L}'(f_v) = 20 \log_{10}\!\left(\frac{\Gamma\,A\,f_0}{2 f_v}\right),
  \label{eq:vibration-phase-noise} 
\end{equation}
where $A$ is the acceleration amplitude and $f_0 = \SI{10}{MHz}$ is the carrier; the $1/f_v$
dependence places the largest contribution at the low frequencies where the balloon's power
concentrates. Folding the BVEXTracker accelerations through Equation~\ref{eq:vibration-phase-noise}
gives, on the strongest ($a_x$) axis, $\mathcal{L}'(1\,\mathrm{Hz}) \approx -142$\,dBc/Hz and
$\mathcal{L}'(10\,\mathrm{Hz}) \approx -162$\,dBc/Hz. These estimates lie about $34$\,dB and $26$\,dB below
the oscillator's static datasheet limits of $-108$ and $-136$\,dBc/Hz, respectively. The estimated
vibration contribution to the OCXO phase noise is therefore well below the intrinsic oscillator phase noise
across the pendulation band. Figure~\ref{fig:ocxo-vibration} shows the
measured acceleration environment.

\section{SYSTEM INTEGRATION AND OPERATIONAL SOFTWARE}
\label{sec:integration}

The BVEX backend and timing system are operated by a network of computers housed on the
gondola and controlled from the ground over an S-band telemetry link. Two flight computers run
the BVEX experiment: the backend and data-acquisition computer (Sagittarius), which commands the RFSoC and logs
the housekeeping and timing data, while the pointing computer (Ophiuchus) controls the telescope elevation and logs the attitude and position of the telescope \cite{thiel2026bvex}. A third computer (Aquila) captures the
100~GbE VLBI stream described in Section~\ref{subsec:storage}. The pointing, thermal, and
power-control software are outside the scope of this paper; here we describe the software that
operates the backend and the timing chain.

\subsection{Backend control and acquisition}
\label{subsec:bcp}

The backend is operated by the \texttt{BCP}, the flight software running on the control
computers. \texttt{BCP} can be commanded to select an observing mode by the user. The code configures the RFSoC by issuing a command to a lightweight daemon running on the RFSoC's
processor, which loads the requested bitstream (spectrometer or VLBI), so a change of mode is a
single command from the ground (Section~\ref{subsec:casper}). Another command,
\texttt{rfsoc\_configure\_ocxo}, directs the same daemon to program the board's LMK jitter cleaner
and LMX synthesizers from the prepared register files, which locks the converter clocks to the
external \SI{10}{MHz} reference (Section~\ref{subsec:rfsoc}). The daemon reports the result back
to \texttt{BCP}, and the same clock configuration serves both firmware modes.

In spectrometer mode, a process on the RFSoC reads the accumulated spectrum from the eight parallel
block-RAM banks through the \texttt{casperfpga} interface at about \SI{3.7}{Hz}. It extracts the 167-channel
window around \SI{22.235}{GHz}, subtracts a baseline, and writes the reduced spectrum into a
shared-memory segment. A server on the control computer reads that segment and forwards the spectrum
to the ground, where pointing the telescope at a massive star-forming region with bright water maser emission confirms that the pointing model is correct. The full-resolution spectra are written to solid-state drives (SSDs) on Sagittarius
(Section~\ref{subsec:spectrometer}). In VLBI mode, a controller process on the storage computer
manages the capture and takes start and stop commands from the control computer. On start, it
enables the RFSoC's 100~GbE output, receives the two-bit packet stream, and writes it to the NVMe
array in time-rotated files with accompanying metadata. The capture process inspects the two timing counters
as packets arrive, tracks changes in the seconds field, and flags the cycle field when it does not
reset at the 1~PPS boundary (Section~\ref{subsec:vdif}). These diagnostics characterized the
timestamp generator's behavior before the flight as discussed in Section~\ref{subsec:drift}.

\subsection{Timing, housekeeping, and autonomy}
\label{subsec:autonomy}

The control computer reads the timing chain directly. It logs the TICC drift measurements and the
GNSS receiver's position, velocity, and time over serial links, rotating the log files at fixed
intervals so that no single file grows without bound. These streams, together with the backend
status, feed a single telemetry aggregator on the control computer. The VLBI data are far too large
to downlink and are never sent to the ground; instead the backend reports only compact metadata which includes the
capture stage, the running packet count, the current data-file size, the space remaining on each
storage drive, and the storage computer's CPU temperature. These values refresh
approximately every \SI{10}{s} and let an operator confirm that recording is active and that the
disks have enough free space for the planned observations.

The control computer also serves as the payload's time server. A Network Time Protocol (NTP)
service, implemented with \texttt{chrony}, takes its time from the GNSS receiver's serial messages and refines it with the GPS PPS. This sharpens the synchronization from the
millisecond level of network time alone to the microsecond level. The other payload computers
synchronize to this server over the BVEX network, so logs, sensor records, and housekeeping
entries carry consistent timestamps across the experiment. The VLBI data, in contrast, take their timestamps from the OCXO-driven timing chain discussed in
Section~\ref{sec:timing}.

The BVEX experiment is built to be able to run independent of operator control. Each subsystem runs as a supervised system service that restarts
automatically if it exits, and the two flight computers (i.e., Sagittarius and Ophiuchus, not Aquila) exchange a readiness handshake at start-up so
the backend does not begin acquisition before the rest of the experiment is initialized and operational. Data acquisition and
logging processes degrade gracefully: a lost connection to the RFSoC or the storage computer is
retried, not treated as fatal, and the timing and housekeeping logs continue independently of
the science data capture. This autonomy is important for a balloon-borne experiment, where telemetry availability is intermittent and can have significant latency, meaning that an
operator cannot intervene at every step.




\begin{figure}[tb]
  \centering
  \includegraphics[width=\textwidth]{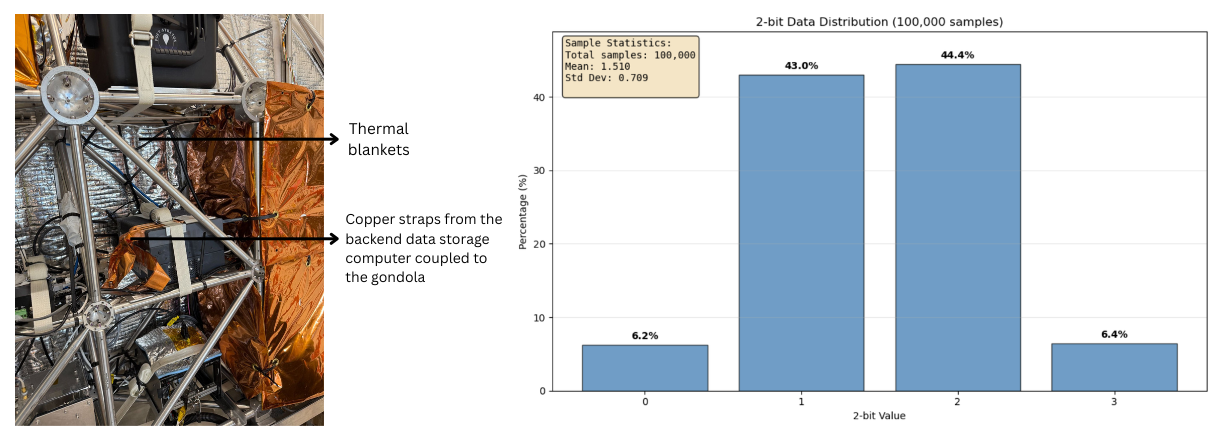}
  \caption{Left: the backend data storage computer mounted in the BVEX gondola for the August
  2025 flight, wrapped in multilayer thermal blankets, with copper straps coupling the computer to
  the gondola frame as the conductive heat path. Right: measured two-bit VLBI data statistics from
  ground testing. In a 100{,}000-sample snapshot the four levels hold 6.2\%, 43.0\%, 44.4\%, and
  6.4\% of the samples; the outer-level occupancy implies the fixed $\pm1000$ thresholds sat near
  $\pm1.5\sigma$ of the noise at this input level (Section~\ref{subsec:datapath-validation}).}
  \label{fig:payload-integration}
  \label{fig:2bit-stats}
\end{figure}

\section{RESULTS AND PERFORMANCE}
\label{sec:results}

The August 2025 flight ended before reaching float altitude, so the system was not exercised on
sky during the flight. Its performance therefore rests on ground and laboratory validation, which we summarize here
together with the in-flight behavior observed during the truncated ascent. We report the data-path
validation, the spectrometer sensitivity confirmed by a water-maser detection, and the timing-chain
performance, and then describe the flight itself.

\subsection{Backend Pre-flight Tests and Validation}
\label{subsec:datapath-validation}

A two-bit VLBI data capture during our pre-flight testing in Timmins showed
a symmetric, stable four-level distribution: in a representative 100{,}000-sample snapshot the levels
held 6.2\%, 43.0\%, 44.4\%, and 6.4\% of the samples (mean symbol value 1.51, against 1.5 for a
balanced quantizer), steady across the record (Figure~\ref{fig:2bit-stats}). The near symmetry shows
that the positive and negative quantizer states were similarly populated. The outer levels held only
about 6\% of the samples each,
below the $\sim$16\% expected at the SNR optimum, so at this test's input level the fixed $\pm1000$
thresholds sat near $\pm1.5\sigma$ of the noise rather than the optimal $\pm1\sigma$
(Section~\ref{subsec:requant}). This is a direct measurement of the fixed-threshold behavior and the
motivation for the adaptive, power-tracking thresholds planned for BVEX~2.0 (Section~\ref{subsec:bvex2-firmware}).

Ground tests showed that the timing chain met its requirements, as detailed in Section~\ref{sec:timing}.
The free-running OCXO drifted against GPS at \num{-49.4}\,ns/s in this measurement, with residuals of \SI{2.21}{ns}
about the linear trend; this drift rate is specific to that run and is not necessarily constant. Its Allan deviation at \SI{1}{s} was below \num{4e-13} across repeated tests. The thermoelectric controller held the OCXO inside the pressure vessel to within
\SI{0.1}{\celsius} of its \SI{25}{\celsius} set point, and the counter resolved the drift well
enough for the planned post-processing correction.

\subsection{Pre-flight single-dish BVEX observations of the W49N water-maser}
\label{subsec:w49n-detection}

We validated the end-to-end spectrometer sensitivity by detecting the \SI{22.235}{GHz} water maser in
W49N with the \SI{0.9}{m} BVEX telescope. W49N is a bright, highly variable Galactic water maser. One monitored
feature rose from $8.7$\,kJy to $84$\,kJy during a 2017 flare \cite{volvach2019w49n}. Its high
line flux makes W49N useful for checking both pointing and the receiver-plus-backend chain for a small radio telescope. For a
single-polarization total-power spectrometer the radiometer equation gives a per-channel noise of
$\sigma_T = T_\mathrm{sys}/\sqrt{\Delta\nu\,\tau}$, where $T_\mathrm{sys}$ is the system temperature,
$\Delta\nu$ the channel width, and $\tau$ the integration time \cite{tms2017interferometry}. With the
room-temperature receiver ($T_\mathrm{sys} \approx \SI{400}{K}$) and the \SI{120}{kHz} channel, and
accounting for the dilution of the $\sim$\SI{37}{kHz} ($\approx 0.5$\,km/s) maser line within that wider channel, the
radiometer model predicts a $5\sigma$ detection time of order ten minutes for a $5$\,kJy feature. This estimate is conservative: brighter features, which W49N commonly reaches (tens of kJy during flares), are detected in a few minutes, consistent with the line appearing in 60-second integrations of our data.

\begin{figure}[tb]
  \centering
  \includegraphics[width=0.8\textwidth]{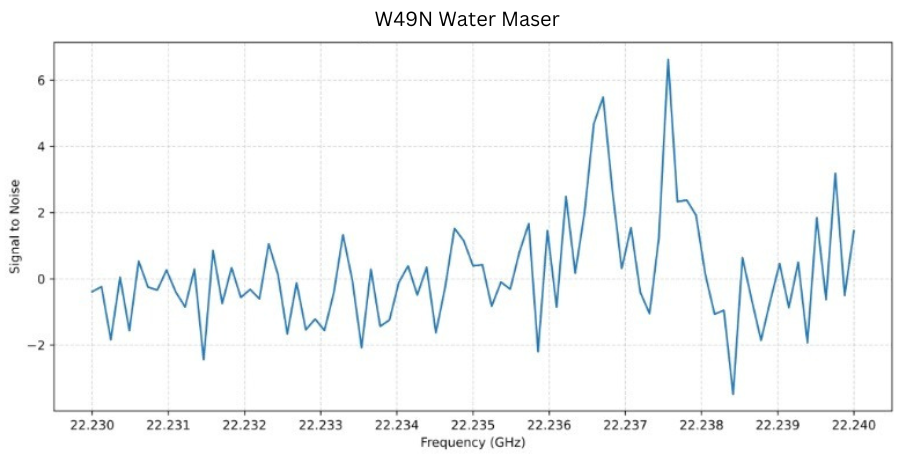}
  \caption{Detection of the \SI{22.235}{GHz} W49N water maser with the BVEX \SI{0.9}{m} dish,
  shown as signal-to-noise ratio versus frequency and
  recorded in eight minutes of integration at \SI{120}{kHz} resolution. The detection validates the
  end-to-end sensitivity of the antenna, receiver, and spectrometer, consistent with the
  radiometer equation for the measured system temperature and channel width once dilution of the
  $\sim$\SI{37}{kHz} line is included.}
  \label{fig:w49n-detection}
\end{figure}

We recorded the maser in an eight-minute integration. This ground measurement tests the antenna, receiver, and spectrometer together, and is shown in Figure~\ref{fig:w49n-detection}. The dominant limit on
detection speed is the system temperature, which is mostly set by the room temperature BVEX receiver, together with the channel width relative
to the maser line. The receiver improvements planned for the 2027 BVEX flight that will lower this system temperature are described in the
companion paper \cite{thiel2026bvex}.

\subsection{The August 2025 BVEX flight}
\label{subsec:flight}

BVEX was launched on its first flight on August 29, 2025, from the CSA STRATOS base in Timmins, Ontario. A leak in the
balloon envelope reduced the ascent rate during the climb, and despite the release of ballast, the balloon failed to reach the target altitude (35 km) and could not ascend above 16.5 km. It remained trapped in the tropopause, drifting south,
until the flight was terminated after about three hours. The payload was recovered, but no science
observations were obtained because the balloon gondola azimuth, and therefore the telescope azimuth, could not be controlled at such low altitudes.

The backend and timing systems remained powered through the ascent and returned housekeeping and
thermal telemetry, but the payload did not record a VLBI observation. The most informative product
of the flight is its temperature record.
Trapped near the tropopause, the payload spent hours in ambient air at roughly
\SIrange{-60}{-50}{\celsius}. The coldest backend sensor, on the RFSoC chassis, fell to about
\SI{-35}{\celsius}, below the \SI{-30}{\celsius} nighttime stratospheric temperature typical of the latitude of Timmins, while the storage-computer chassis and the NIC stayed near \SIrange{-5}{0}{\celsius}. The flight never
reached the low-pressure float environment for which the thermal modelling was performed; the
temperature record and the recovered hardware informed the redesign described in
Section~\ref{sec:bvex2}. The full flight trajectory is presented in the
companion paper \cite{thiel2026bvex}.


\section{LESSONS LEARNED AND NEXT-GENERATION BACKEND}
\label{sec:bvex2}

The reflight is planned for 2027. The preferred launch is an overnight zero-pressure-balloon flight
from Palmas, Brazil, with Timmins, Ontario as a fallback. That profile gives about four hours of
nighttime observations, enough to attempt the first balloon-to-ground VLBI fringes with the
validated backend and timing system.
 Due to the truncated flight during our 2025 Timmins campaign, a second flight is required to meet the objectives of the BVEX experiment. For the second version of the experiment (BVEX 2.0), we are planning upgrades to the backend system, and the subsequent hardware inspection identified changes for the redesigned
BVEX 2.0 backend. We will retain the original signal-processing architecture while improving the thermal
management of the RFSoC and the data storage computer. The backend will use a new data storage computer design, and we will improve the RFSoC firmware that outputs VLBI data to make the packets more compatible with \texttt{DiFX}~\cite{deller2011difx}.

\subsection{Lessons from the 2025 flight}
\label{subsec:lessons}

The clearest lesson concerns thermal design. At float altitudes of 35 km, low pressure makes convective cooling weak,
and the multilayer insulation that protects most of the gondola limits radiative heat rejection as shown in Figure~\ref{fig:2bit-stats}; the
electronics therefore require a controlled conductive path to the gondola structure. The 2025 backend used a copper strap to transport heat from the RFSoC chip to its electronics box chassis. The storage computer, however, used a commercial-off-the-shelf industrial chassis fitted with an add-in
100~GbE NIC and NVMe drives. The enclosed layout of this computer coupled poorly to the gondola and constrained the
internal heat dissipation. Because the flight never reached stratospheric float, the thermal design was never
exercised in its intended low-pressure environment. 
The flight also motivated two RFSoC firmware changes for BVEX 2.0, a PPS-synchronous reset of the sub-second cycle counter and adaptive two-bit thresholds for the VLBI data capture, which we describe in Section~\ref{subsec:bvex2-firmware}. 
\subsection{Next-generation backend: compute, storage, and capture}
\label{subsec:bvex2-backend}

The BVEX 2.0 design replaces the commercial storage computer with a custom Mini-ITX system built
around an AMD Ryzen 5~7600X processor and 32\,GB of memory. A custom, mechanically modeled chassis will conduct heat from the
hot components (the processor, the 100~GbE NIC, and the NVMe drives) directly into the chassis walls
through dedicated heat sinks. The chassis will in turn be thermally bonded to the gondola structure, so
conduction is the primary heat path by design. The design will incorporate three \SI{8}{\TB} NVMe drives with a total of \SI{24}{\TB} of storage for VLBI data. 

The 2025 100 GbE NIC operated through the NVIDIA/Mellanox Linux driver stack. The acquisition
software used a raw-packet catcher and a Python logging process: the logger received packet batches
over a local interprocess connection, checked the custom header, and wrote binary files using
standard buffered file I/O. It did not use \texttt{O\_DIRECT} (writing directly to storage, bypassing the operating-system page cache), \texttt{recvmmsg} (receiving many packets in a single system call), or a user-space storage path (moving data from the network card to disk without passing through the kernel). 

For BVEX 2.0, we replaced this path with a kernel-bypass capture engine built on \texttt{AF\_XDP} (a
Linux socket that delivers received packets to a user-space buffer while bypassing the kernel network
stack). A small \texttt{XDP}/\texttt{eBPF} program (a sandboxed filter that runs in the network
driver) steers the VLBI flow into this socket, and a writer thread commits the packets to the NVMe
array with \texttt{O\_DIRECT}, which bypasses the operating-system page cache. We also evaluated a simpler
\texttt{recvmmsg}-based receive path and the fully user-space frameworks, the Data Plane Development
Kit (\texttt{DPDK}) for packet reception and the Storage Performance Development Kit (\texttt{SPDK})
for NVMe writes, but the \texttt{AF\_XDP} and \texttt{O\_DIRECT} combination already sustains the
single \SI{8.2}{\Gbps} flow, so the heavier user-space frameworks proved unnecessary. We have
demonstrated this capture path on the bench at the full VLBI data rate without packet loss, and will
harden it for sustained, full-disk recording ahead of the 2027 reflight.

\subsection{RFSoC Firmware and Timing System Upgrades}
\label{subsec:bvex2-firmware}

The planned next-generation firmware implements the two refinements noted above. We will replace the
fixed two-bit thresholds with adaptive ones recomputed from the running signal power, so that they
stay near $\pm1\sigma$ of the noise as the receiver gain and the sky and source conditions
change. We will also replace the custom timestamp fields with the VDIF reference epoch, seconds, and
frame number, with the frame counter reset synchronously on a registered PPS edge. This change will
remove the timestamp correction now required before conversion to standard VDIF. For the timing system,
we plan to upgrade the OCXO's thermal-control stage with a higher-power heater for better temperature
regulation at float, and to rebuild the time-interval counter box around the revised PPS logic.

\section{CONCLUSIONS AND FUTURE WORK}
\label{sec:conclusions}

We have presented the digital backend and precision-timing system for BVEX, a K-band radio telescope that aims to make a first VLBI detection between balloon-borne and ground-based telescopes. The BVEX backend digitizes a \SIrange{2}{4}{GHz} receiver IF on a single RFSoC
4x2 ADC, then either channelizes it for amplitude spectrum observations, or produces a two-bit baseband stream of VLBI data transmitted at about
\SI{8.2}{\Gbps} over 100~GbE. The timing system pairs a free-running Rakon RK409 OCXO
with continuous GPS-referenced drift monitoring, so the recorded samples can be aligned with
ground-array data during the post-flight correlation of the telescope data. This is the first system of its kind, an RFSoC-based VLBI backend
with its own precision-timing chain, to fly on a stratospheric balloon.

Ground and laboratory testing validated the principal backend and timing functions. A detection of the 22 GHz water-maser spectral line toward the massive star-forming region W49N was made in eight minutes, which confirmed the spectrometer sensitivity against predictions based on the radiometer
model. The timing chain was measured to have an Allan deviation below \num{4e-13} at
\SI{1}{s}, with drift monitored at the \SI{60}{ps} level; this meets the coherence error budget for
\SI{22}{GHz} VLBI with considerable margin. The August 2025 CSA STRATOS flight ended before
reaching float after a balloon failure, so no science observations were obtained. The backend and
timing systems returned housekeeping telemetry during the ascent, and the hardware was recovered.

The BVEX team is currently upgrading the experiment for a second flight (BVEX 2.0) to demonstrate balloon-borne VLBI in 2027. The BVEX 2.0 design uses a custom,
conduction-cooled storage computer with \SI{24}{\TB} of NVMe storage. Planned software and firmware
upgrades include an \texttt{AF\_XDP}/\texttt{O\_DIRECT} kernel-bypass NIC-to-NVMe capture path, adaptive requantization for the 2-bit VLBI data, and complete VDIF
timestamp fields with a synchronous PPS reset. With these changes, the 2027
reflight aims for the first balloon-to-ground VLBI fringes. The compact architecture may also be
applicable to other power- and mass-constrained radio platforms.

\acknowledgments

The Balloon-borne VLBI Experiment (BVEX) was funded by the Canadian Space Agency's (CSA) Flights and
Fieldwork for the Advancement of Science and Technology (FAST)'s 2021 AO (21FAQUEA18), with supporting funding from the Queen's Faculty of
Arts Infrastructure Fund,  the Canada Foundation for Innovation (CFI) John R. Evans
Leaders Fund (JELF Project Number 43766), National Science and Engineering Research Council
(NSERC)  Discovery Grant RGPIN/06266-2020, and
the Queen's University Research Initiation
Grant.  The upgraded BVEX2.0 experiment is funded by the CSA's FAST 2025 AO (25FAQUB72). We acknowledge observing support from the Very Long Baseline Array (VLBA) of
the National Radio Astronomy Observatory (NRAO), the Max-Planck-Institut f\"ur Radioastronomie
(MPIfR), and the Effelsberg \SI{100}{m} Radio Telescope. We thank the CASPER community for its
open-source tools and technical support. We particularly thank Mitchell Burnett for assistance with the RFSoC firmware.
We also thank Jens Kauffmann, Assistant Director of MIT Haystack Observatory and the whole Haystack Observatory team, for advice in designing BVEX and support during
ground testing and coordinated observations. We would also like to thank the engineers and scientists National Research Council's Herzberg Astronomy and Astrophysics Research Centre and Dominion Radio Astrophysical Observatory for their advice and for participating in key system design reviews. Finally, we would like to thank the CSA STRATOS and CNES teams for supporting our 2025 Timmins flight campaign.

\bibliography{report} 
\bibliographystyle{spiebib} 

\end{document}